\def\al26{\mbox{$^{26}$\hspace{-0.2em}Al}}          
\def\MeV{\mbox{Me\hspace{-0.1em}V}}                 
\def\deg{\mbox{$^\circ$}}                           
\def\dg{^\circ}                                     
\def\funit{photons cm$^{-2}$ s$^{-1}$}              
\def\phibar{$\bar{\varphi}$}                        
\def\Msol{\mbox{M$_{\odot}$}}                       
\def\etal{{et al.}}                                 
\def\gray{\mbox{$\gamma$-ray}}                      
\def\sigmad{\mbox{$\Sigma$-D}}                      
\def\reference{}

\documentstyle[psfig]{l-aa}

\begin{document}

\thesaurus{06(13.07.2; 02.14.1; 08.19.4; 09.19.2)}

\title{Search for 1.809 \MeV\ Emission of \al26\ from nearby Supernova
Remnants using COMPTEL}

\author{J\"urgen Kn\"odlseder$^{1,2}$, Uwe Oberlack$^2$,
        Roland Diehl$^2$, Wan Chen$^{3,4}$, and Neil Gehrels$^3$}

\institute{$^1$Centre d'Etude Spatiale des Rayonnements, CNRS/UPS, BP 4346,
	           31029 Toulouse Cedex, France\\
	       $^2$Max Planck Institut f\"ur extraterrestrische Physik,
	           Postfach 1603, 85740 Garching, Germany\\
	       $^3$NASA/Goddard Space Flight Center, Code 661, Greenbelt,
	           MD 20771, USA\\
	       $^4$Universities Space Research Association
}

\offprints{J\"urgen Kn\"odlseder (Toulouse)}

\date{Received October 1995; accepted October 1995}

\maketitle
\markboth{J. Kn\"odlseder et al.: Search for 1.809 \MeV\ Emission of
          \al26\ from nearby SNRs using COMPTEL}
	     {J. Kn\"odlseder et al.: Search for 1.809 \MeV\ Emission of
          \al26\ from nearby SNRs using COMPTEL}

\begin{abstract}

We report the negative results of our searches in COMPTEL data for
1.809 \MeV\ gamma-ray line emission from four localized regions which
contain nearby supernova remnants (SNRs).
The upper flux limits (2$\sigma$) are found to be in the range of
$1.4 \times 10^{-5}$ to $2.4 \times 10^{-5}$ photons s$^{-1}$ cm$^{-2}$.
These upper limits do not severely constrain the theoretical \al26\ yields
from individual core collapse supernovae due to large uncertainties in the SNR
distances and the nature of the progenitor stars.

\keywords{gamma-rays: observations -- nucleosynthesis -- supernovae --
          ISM: supernova remnants}
\end{abstract}

\section{Introduction}

One of the outstanding achievements of the Compton Imaging Telescope (COMPTEL)
aboard the {\em Compton Gamma Ray Observatory} has been the first sky map
in the light of the 1.809 \MeV\ \gray\ line which is attributed to
radioactive decay of \al26\ ($\tau=1.04\times10^6$ yr).
The observed emission is clearly of Galactic origin since it is concentrated
on the Galactic plane.
Its distribution along the plane is strikingly lumpy with extended
emission features and `hot-spots' (\cite{rf:drea95a}).
One of these features, situated in the Vela region, is of particular
interest (Diehl \etal\ 1995b).
A recent survey of candidate \al26\ sources (core collapse supernovae (SNe),
Wolf-Rayet (WR) stars, asymptotic giant-branch (AGB) stars, and O-Ne-Mg
novae) in this region of the sky by Oberlack \etal\ (1994) identified the
Vela supernova remnant (SNR) as most likely source of the emission if its
progenitor star was massive ($\sim35$ \Msol) and its distance is $\le350$ pc.
The distance constraint is in line with recent X-ray based distance
estimates of 350, 400-600, and 125-160 pc (Aschenbach 1993, Aschenbach
et al. 1995, Becker 1995), respectively, all pointing towards distances
below the canonical 500 pc.
These findings raise the question if there are other SNRs detectable
by COMPTEL as \al26\ sources.
{}From the 1.8 \MeV\ \gray\ line sensitivity of $\sim10^{-5}$ \funit\
(for an observation of $10^6$ seconds) and an optimistic \al26\ supernova yield
of $3\times10^{-4}$ \Msol\ for type II SNe (Hoffman \etal\ 1995) we
estimate that SNe could be detectable by COMPTEL up to
distances of $\sim600$ pc.
In this paper we report on our attempt to identify probable candidate
SNRs and on the search for their 1.809 \MeV\ \gray\ line emission.

\section{Supernova Remnants as \al26\ source tracer}

\subsection{SNR types}

While SNRs are assumed to be produced by all types of SNe, only core
collapse events (type II and Ib) are expected to release significant
amounts of \al26\ (Leising 1994).
Unfortunately, there is no certain, direct way to determine which of
the Galactic SNRs are associated with core collapse events.
For the historical SNe, knowledge of the light curve sheds some light
on their type, but the identification is not unambigous (Doggett \&
Branch 1985).
The presence of a neutron star within a SNR would identify it as core
collapse remnant, but there are only 10 known Galactic SNR-pulsar
associations (\cite{rf:cp93}).
All of them except the Vela SNR, however, are at distances greater
than 1.8 kpc, much too far to be detectable by COMPTEL (see above).
Recently, evidence is growing that oxygen-rich and plerionic-composite
SNRs may come from type Ib and type II SNe respectively
(\cite{rf:vdB88}, \cite{rf:ws88}, \cite{rf:sd95}).
However, most of the Galactic SNRs don't fall in these classes.

\subsection{SNR distances}

\begin{figure}
 \setlength{\unitlength}{1cm}
 \begin{minipage}[t]{8.8 cm}        
  \begin{picture}(8.8,6)          
   \framebox(8.8,6){
    \psfig{figure=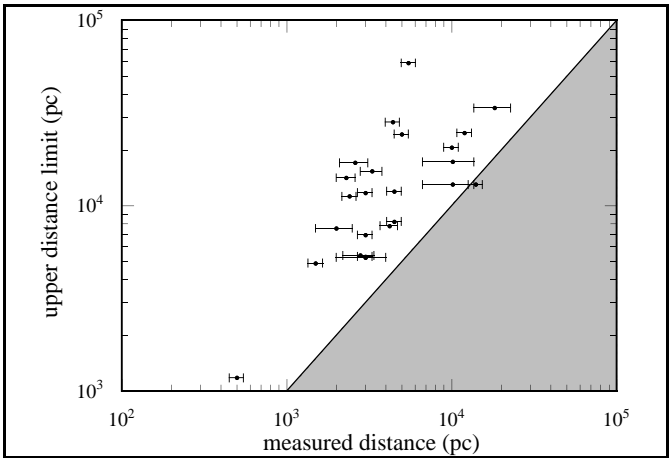,height=6cm}
     }
  \end{picture}
  \caption{\label{fig:indicator}
	   Upper distance limits derived by Berkhuijsen's (1986)
	   \sigmad\ relation for 23 SNRs with good or reasonable
	   distance measurements. The grey-shaded area marks the range
	   where the \sigmad\ relation contradicts the measurements.}
 \end{minipage}
\end{figure}

The key element in our candidate source selection is the distance of
the remnant.
Of the 194 SNRs known in our Galaxy (\cite{rf:gda95}), only 23 have good
or reasonable distance estimates (\cite{rf:gda91}).
{}From those, only the Vela SNR is close enough to be detectable by COMPTEL.
For the other SNRs, only rough or no distance estimates are available which
makes the direct determination of the nearest \al26\ source candidates
unfeasible.
Therefore, we used the empirical radio surface brightness - diameter
relation (\sigmad\ relation) to identify the closest SNRs
(e.g. \cite{rf:berk86}).
Green (1984, 1991) pointed out that the \sigmad\ relation cannot provide
accurate distances to individual SNRs, but Berkhuijsen (1986) showed
that it sets reliable upper distance limits:
using $\Sigma \propto \mbox{D}^{-3.5}$ (\cite{rf:berk86}), none from
23 SNRs are found in the region where the distance measurement contradicts
those upper limits (see Fig. \ref{fig:indicator}, grey-shaded area).

\subsection{Candidates for SNR search}

Using Berkhuijsen's (1986) \sigmad\ relation as distance indicator,
we selected four SNRs as possible candidates for nearby \al26\ sources:
the Cygnus Loop, the Monoceros Nebula, the Lupus Loop and HB 21.
Selecting the SNRs with the lowest distance limits obviously chooses
nearby objects, but not necessarily all of them.
There could be nearby SNRs which are not extended or not very powerful and
therefore have large distance limits.
None of the selected remnants is in Green's (1984, 1991) lists of good or
reasonable distance determinations, but crude estimates are available for
all of them.
We compiled the relevant properties of the four candidates in Table
\ref{tb:snrs} along with those of the Vela SNR.

The Cygnus Loop and HB 21 were probably created by core collapse events.
X-ray observations with EINSTEIN provide evidence for cavities devoided
of clouds prior to the explosions, probably created by the stellar winds
of the progenitor stars (\cite{rf:charles85}, \cite{rf:leahy87}).
Since strong stellar winds are only expected for massive stars, the
SNe must have been of type II or Ib.

Also the Monoceros Nebula was probably created by a core collapse event.
Odegard (1986) summarizes observational evidence which indicates that
the Monoceros Nebula is associated with the Mon OB2 stellar association
and the Rosette Nebula.
Since low-mass stars are generally not very evolved in OB associations,
the occurence of type Ia SNe is rather improbable.
Hence if a SNR is located in an OB associations it is likely that it arose
from the explosion of a massive star.

We found no indication of SN type for the Lupus Loop in the literature.
However, the (fairly uncertain) distance estimate of 1.2 kpc
(\cite{rf:leahy91}) and its high Galactic latitude of $b$=15.97\deg\ places
it 330 pc above the Galactic plane.
This height is more typical for the old stellar population than for young
massive objects which would favour the relation of the SNR to a type Ia
event.
Nevertheless, we included this object in our 1.8 \MeV\ emission search.

\begin{table*}
\caption[]{Properties of nearby supernova remnants (Green 1984, 1991).
           The upper distance
           limits were derived using the \sigmad\ relation
           $\Sigma_{\rm{1GHz}}=2.51 \times 10^{-14}$ D$^{-3.5}$
           W Hz$^{-1}$ m$^{-2}$ sr$^{-1}$ of Berkhuijsen (1986).}
\begin{flushleft}
\begin{tabular}{lrrcccccc}
\noalign{\smallskip}
\hline
\noalign{\smallskip}
Name & \multicolumn{1}{c}{$l$} & \multicolumn{1}{c}{$b$}
& $F_{\rm{1GHz}}$ & $\Sigma_{\rm{1GHz}}$ & Radio Size D & Distance &
  Upper distance limit \\
& (deg) & (deg) & (Jy) & (W Hz$^{-1}$ m$^{-2}$ sr$^{-1}$) & (arcmin) & (kpc) &
  (kpc) \\ \hline
 Vela SNR    & 263.9  & -3.3 & 1750 & $4.1 \times 10^{-21}$ & 255 & $<0.5$
  & 1.2  \\
  & & &  \\
 Cygnus Loop & 74.01  & -8.56 & 210 & $6.0 \times 10^{-22}$ & 195 & $0.77^a$
  & 2.4  \\
 HB 21       & 88.86  &  4.80 & 220 & $2.3 \times 10^{-21}$ & 105 & $0.8^b$
  & 3.1  \\
 Mon Nebula  & 204.96 &  0.45 & 160 & $5.0 \times 10^{-22}$ & 220 & $1.6^c$
  & 2.5  \\
 Lupus Loop  & 329.67 & 15.97 & 350 & $1.6 \times 10^{-21}$ & 180 & $1.2^d$
  & 2.2  \\
\noalign{\smallskip}
\hline
\noalign{\smallskip}
\end{tabular}
\\
{\footnotesize
$^a$ from a combination of optical proper motion and radial velocity
     observations (\cite{rf:minkowski58}). \\
$^b$ from a possible association with Cyg OB7 (\cite{rf:tatematsu90}). \\
$^c$ from a possible association with Mon OB2 and the Rosette nebula
     (\cite{rf:odegard86}). \\
$^d$ from a model-fit to X-ray data, assuming a SN explosion energy of
     $E_0=10^{51}$ erg (\cite{rf:leahy91}).
}
\label{tb:snrs}
\end{flushleft}
\end{table*}

\section{Observations and Data Analysis}

We used COMPTEL data collected between May 1991 and October 1994 for
the 1.8 \MeV\ \gray\ line search.
In summary, $\sim100$ different pointings (observation periods) were
combined, yielding a 1.8 \MeV\ \gray\ line sensitivity of
$\sim1.5\times10^{-5}$ \funit\ ($2\sigma$) for all candidate sources.
Measured \gray\ photons with energies between 1.7-1.9 \MeV\ were binned
in a three-dimensional data-space, spanned by the scatter direction
($\chi,\psi$)
(defined by the photon's interaction locations in the two detector layers
of the  instrument), and the Compton scatter angle \phibar\ (defined by
the energy deposits in these two layers; see Sch\"onfelder \etal\ (1993)
for a detailed description of the instrument).
For each SNR search, a data-space enclosing $\pm50\dg$ around the SNR
position was employed.
The data-space distribution of instrumental background was derived from
measurements at adjacent energy bands (\cite{rf:knoedl95}).
The search for 1.8 \MeV\ emission was carried out by convolving models of
the SNRs into the data-space which were fitted along with the
background model to the data.
The maximum-likelihood technique (\cite{rf:deboer92}) was applied to
determine flux, flux error, and detection significance for all models.
Upper limits of 1.8 \MeV\ emission were determined by variation of the SNR
source flux to obtain a likelihood-ratio corresponding to $2\sigma$ limits
above the best-fit (or above zero flux in cases where the best fit gave a
negative source flux).

\section{Results}

Since it is not clear how the \al26\ is distributed within the SNRs, we
tested two extreme hypotheses for each candidate:
(1) all \al26\ is confined to the center of the remnant (point source
hypothesis), and
(2) the \al26\ is distributed homogeneous over the SNR (extended source
hypothesis).
However, we don't expect large differences in the results from both
hypotheses because the sizes of the SNRs are of the same order as
COMPTEL's angular resolution of 3.8\deg\ (FWHM).

We did not detect significant 1.8 \MeV\ emission from any of the four SNR
positions.
The upper flux limits ($2\sigma$) for all remnants are listed in Table
\ref{tb:upperlimit} for the point source and extended source hypotheses.
The last column quotes the assumed diameters for the extended source
hypotheses.
In general, the extended source models yield somewhat higher limits
than the point source models, particularly for the Monoceros Nebula
which lies in the vincinity of an insignificant 1.8 \MeV\ emission feature.
The highest limits for both the point source and the extended source
hypotheses were obtained for HB 21.
Formally, HB 21 yields a small signal, but since this SNR is embedded in
an extended 1.8 \MeV\ emission feature along the direction of Cygnus,
the signal can possibly be attributed to the local spiral arm
(for a discussion of 1.8 \MeV\ emission from the Cygnus region
see del Rio \etal\ (1994)).
We found no evidence for 1.8 \MeV\ emission from HB 21 on top of the
extended emission.

\begin{table}
\caption[]{COMPTEL upper limits ($2\sigma$) for 1.8 \MeV\ emission
	   from four nearby SNRs.}
\begin{flushleft}
\begin{tabular}{lccc}
\noalign{\smallskip}
\hline
\noalign{\smallskip}
Name & Point source & \multicolumn{2}{c}{Extended source} \\
\cline{3-4}
& upper limit & upper limit & diameter \\
& \multicolumn{2}{c}{(\funit)} & (deg)    \\
\hline
 Cygnus Loop & $1.5 \times 10^{-5}$ & $1.5 \times 10^{-5}$ & 4.0 \\
 HB 21       & $2.3 \times 10^{-5}$ & $2.4 \times 10^{-5}$ & 2.0 \\
 Mon Nebula  & $1.6 \times 10^{-5}$ & $2.2 \times 10^{-5}$ & 4.0 \\
 Lupus Loop  & $1.4 \times 10^{-5}$ & $1.6 \times 10^{-5}$ & 3.0 \\
\noalign{\smallskip}
\hline
\noalign{\smallskip}
\end{tabular}
\label{tb:upperlimit}
\end{flushleft}
\end{table}

\section{Discussion}

The two main parameters which affect the expected 1.8 \MeV\ flux from a
SNR are its distance and the amount of \al26\ which was ejected during the
SN explosion.
For type II events, recent nucleosynthesis calculations are available
covering initial progenitor star masses from 11-40 \Msol\ (\cite{rf:ww93};
\cite{rf:hoff95}).
For type Ib events no consistent stellar evolution calculations including
\al26\ nucleosynthesis were followed up to the final SN explosion yet, but
Woosley \etal\ (1993) estimate `a few times $10^{-5}$ \Msol\ of \al26'.
However, if type Ib SNe come from explosions of single WR stars (Woosley \etal\
1995), substantial amounts of the \al26\ which was ejected during the WR
phase (Langer \etal\ 1995) and which dominates the explosion yield
should still be alive, since the time between the onset of the WR
phase and the death of the star is shorter than the \al26\ lifetime.

\begin{figure}
 \setlength{\unitlength}{1cm}
 \begin{minipage}[t]{8.8cm}            
  \begin{picture}(8.8,6)               
    \framebox(8.8,6){
      \psfig{figure=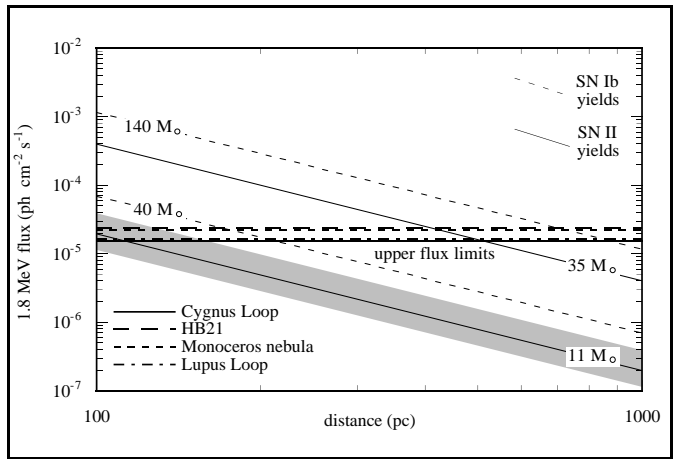,width=8.8cm}
       }
  \end{picture}
  \caption{\label{fig:discussion}
	   Expected SNR 1.809 \MeV\ flux versus remnant distance for type II
	   (Weaver \& Woosley 1993; Hoffman \etal\ 1995) and type Ib SNe
	   (Langer \etal\ 1995). The grey area indicates the range of
	   type II yields from variation of the physical parameters of models
	   for 20 \Msol\ progenitors. The COMPTEL upper flux limits are drawn
	   as horizontal lines. Models which lie above these lines are
	   excluded by the measurement.}
 \end{minipage}
\end{figure}

The wide range of possible \al26\ yields for different SN models does not
allow to put severe constraints on the SNR parameters
(see Fig. \ref{fig:discussion}).
Assuming that the observed SNRs originate from type II events, the upper
flux limits correspond to minimum remnant distances between
90 and 500 pc, depending on the initial mass of the progenitor star.
These lower limits are consistent with the distance estimates for the four
SNRs which range from 770 pc for the Cygnus Loop to 1.6 kpc for the Monoceros
Nebula.
For type Ib SN the lower distance limits are somewhat higher, reaching up
to 900 pc for an initial progenitor star mass of 140 \Msol.
Thus, if the SNR distance estimates are reliable, our 1.8 \MeV\ upper
limits exclude such extremly massive progenitor stars for the Cygnus Loop
and HB21.

\section{Conclusions}

Observations of four nearby supernova remnants with COMPTEL do not provide
evidence for 1.8 \MeV\ line emission from radioactive decay of \al26.
We obtain upper flux limits for the Cygnus Loop, HB 21, the Monoceros Nebula,
and the Lupus Loop.
Our limits are the most stringent to date, yet the large uncertainties in
SNR distances and the nature of the progenitor stars do not allow to put
severe constraints on \al26\ yields from individual supernovae.
Based on current nucleosynthesis models, we found lower distance limits
which are consistent with other distance estimates for the investigated
SNRs.

\begin{acknowledgements}
The COMPTEL project is supported by the German government through
DARA grant 50 QV 90968, by NASA under contract NAS5-26645, and by
the Netherlands Organisation for Scientific Research NWO.
\end{acknowledgements}


\end{document}